\documentclass{article}
\PassOptionsToPackage{numbers, compress}{natbib}
\usepackage[preprint]{neurips_2025}
\usepackage{fix-cm}

\usepackage{xcolor}         % colors
\usepackage{graphicx}
\usepackage{xspace}
\definecolor{linkColor}{rgb}{0.2,0.4,0.6}
\usepackage[utf8]{inputenc} % allow utf-8 input
\usepackage[T1]{fontenc}    % use 8-bit T1 fonts
\usepackage[colorlinks=true,linkcolor=linkColor,citecolor=linkColor,filecolor=linkColor,urlcolor=linkColor]{hyperref}       % hyperlinks
\usepackage{url}            % simple URL typesetting
\usepackage{booktabs}       % professional-quality tables
\usepackage{amsfonts}       % blackboard math symbols
\usepackage{stmaryrd}       % for extra math symbols
\usepackage{nicefrac}       % compact symbols for 1/2, etc.
\usepackage{microtype}      % microtypography

\usepackage{graphicx}
\usepackage{arydshln}
\usepackage{booktabs}
\usepackage{multirow}
\usepackage{caption}
\usepackage{subcaption}
\usepackage{makecell}
\usepackage{epigraph}
\usepackage{booktabs} 

\usepackage{CJKutf8}
\usepackage[utf8]{inputenc} % allow utf-8 input
\usepackage[T1]{fontenc}    % use 8-bit T1 fonts
\usepackage{hyperref}       % hyperlinks
\usepackage{url}            % simple URL typesetting
\usepackage{booktabs}       % professional-quality tables
\usepackage{amsfonts}       % blackboard math symbols
\usepackage{nicefrac}       % compact symbols for 1/2, etc.
\usepackage{microtype}      % microtypography
\usepackage{xcolor}         % colors
\usepackage{markdown}

\usepackage{CJKutf8}
\usepackage{graphicx}
\usepackage{booktabs} % for professional tables
\usepackage{amsmath,amsfonts}
\usepackage{algorithmic}
\usepackage{algorithm}
\usepackage{array}
\usepackage{float}
\usepackage{xcolor}
\usepackage{multirow}

\usepackage{tabularx}   % 添加到导言区
\usepackage{booktabs}   % 保证 \toprule 等能正常使用
\usepackage{url}        % 保证 \url 能使用
\usepackage{ragged2e}   % 用于 \raggedright
\usepackage{geometry}

\usepackage{xcolor}
\usepackage{tcolorbox}

\usepackage{graphicx}
\usepackage{xcolor}
\usepackage{graphicx}
\newtcolorbox{markdownBox}{
    colback=gray!20,
    colframe=black,
    breakable,
    arc=3mm,
    title={},
    toptitle=0mm,
    bottomtitle=0mm,
    colbacktitle=gray!20,
    coltitle=black,
    fonttitle=\bfseries,
}

\usepackage{multirow}

\definecolor{abstractbg}{gray}{0.95}
\makeatletter
\renewenvironment{abstract}{%
  \vskip 0.1in
  \begin{tcolorbox}[
    colback=abstractbg, colframe=abstractbg,
    arc=3mm, boxrule=0pt,
    left=6mm, right=6mm, top=4mm, bottom=4mm
  ]
}{%
  \end{tcolorbox}
}
\makeatother

\usepackage{csquotes}

\RequirePackage{algorithm}
\RequirePackage{algorithmic}

\usepackage{amsmath,amsfonts,bm}

\def\eqref#1{equation~\ref{#1}}
\def\1{\bm{1}}

\DeclareMathAlphabet{\mathsfit}{\encodingdefault}{\sfdefault}{m}{sl}
\SetMathAlphabet{\mathsfit}{bold}{\encodingdefault}{\sfdefault}{bx}{n}

\newcommand\our{\textsc{VibeVoice-ASR}}
\newcommand\ourtts{\textsc{VibeVoice}}
\newcommand\sigvae{$\sigma$-VAE}

\newcommand{\huggingface}{\raisebox{-1.5pt}{\includegraphics[height=1.05em]{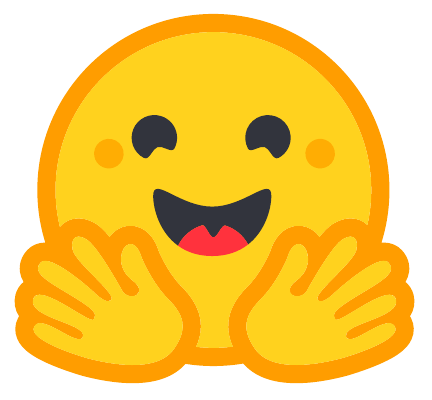}}\xspace}
\newcommand{\github}{\raisebox{-1.5pt}{\includegraphics[height=1.05em]{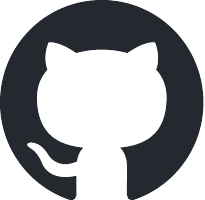}}\xspace}
\newcommand{\microphone}{{\raisebox{-1.5pt}{\includegraphics[height=1.05em]{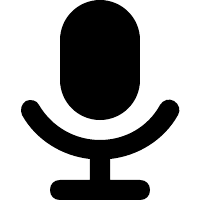}}}\xspace}

\title{\our{} Technical Report
}

\author{
Zhiliang Peng\thanks{~Core contributors. $\diamond$ Contact person: \href{mailto:fuwei@microsoft.com}{fuwei@microsoft.com}.},~~Jianwei Yu\footnotemark[1],~~Yaoyao Chang\footnotemark[1],~~Zilong Wang\footnotemark[1],~~Li Dong\footnotemark[1] \\ 
~\bf Yingbo Hao, Yujie Tu, Chenyu Yang, Wenhui Wang, Songchen Xu, Yutao Sun \\
~\bf Hangbo Bao, Weijiang Xu, Yi Zhu, Zehua Wang, Ting Song, Yan Xia, Zewen Chi \\
~\bf Shaohan Huang, Liang Wang, Chuang Ding, Shuai Wang, Xie Chen, Furu Wei$^{\diamond}$ \\
Microsoft Research \\
~{\href{https://aka.ms/GeneralAI}{https://aka.ms/GeneralAI}}
}

\begin{document}

\maketitle

%\footnotetext{\diamond Contact person: \href{mailto:fuwei@microsoft.com}{fuwei@microsoft.com}}
% \footnotetext*{$\diamond$ Contact person: \href{mailto:fuwei@microsoft.com}{fuwei@microsoft.com}}

% \begin{center}
% \href{https://aka.ms/GeneralAI}{GeneralAI Group}, Microsoft Research
% \end{center}

\begin{abstract}

This report presents \our{}, a general-purpose speech understanding framework built upon \ourtts{}~\cite{vibevoice}, designed to address the persistent challenges of context fragmentation and multi-speaker complexity in long-form audio (e.g., meetings, podcasts) that remain despite recent advancements in short-form speech recognition. Unlike traditional pipelined approaches that rely on audio chunking, \our{} supports single-pass processing for up to 60 minutes of audio. 
It unifies Automatic Speech Recognition, Speaker Diarization, and Timestamping into a single end-to-end generation task. 
In addition, \our{} supports over 50 languages, requires no explicit language setting, and natively handles code-switching within and across utterances.
Furthermore, we introduce a prompt-based context injection mechanism that allows users to supply customized conetxt, significantly improving accuracy on domain-specific terminology and polyphonic character disambiguation.

\end{abstract}

% \begin{table}[H]
% \centering
% \begin{tabular}{@{}r@{\hspace{2pt}}l@{}} % 调整为 2pt
% % \house & \textbf{Project Page}: \href{https://microsoft.github.io/VibeVoice}{\texttt{aka.ms/VibeVoice}}\\
% \github & \textbf{Code}: \href{https://github.com/microsoft/VibeVoice}{\texttt{github.com/microsoft/VibeVoice}} \\
% \microphone & \textbf{Demo}: 
% \href{https://aka.ms/VibeVoice-ASR}{\texttt{aka.ms/VibeVoice-ASR}}\\
% \huggingface & \textbf{Hugging Face}: 
% \href{https://huggingface.co/collections/microsoft/vibevoice-68a2ef24a875c44be47b034f}{\texttt{microsoft/VibeVoice}} \\
% \huggingface & \textbf{Transformers release}: 
% \href{https://github.com/huggingface/transformers/releases/tag/v5.3.0}{\texttt{microsoft/VibeVoice}} \\
% \huggingface & \textbf{Microsoft Foundry}: 
% \href{https://huggingface.co/docs/microsoft-azure/foundry/examples/deploy-vibevoice-asr}{\texttt{microsoft/VibeVoice}}\\
% \end{tabular}
% \end{table}

\begin{table}[H]
\centering
\begin{tabular}{@{}c c@{}}
\github 
\textbf{Code}: \href{https://github.com/microsoft/VibeVoice}{\texttt{github.com/microsoft/VibeVoice}} \\

\microphone 
\textbf{Demo}: \href{https://aka.ms/VibeVoice-ASR}{\texttt{aka.ms/VibeVoice-ASR}} \\

\huggingface 
\href{https://huggingface.co/collections/microsoft/vibevoice-68a2ef24a875c44be47b034f}{\textbf{HuggingFace Models}}
\quad
\href{https://github.com/huggingface/transformers/releases/tag/v5.3.0}{\textbf{Transformers Release}}
\quad
\href{https://huggingface.co/docs/microsoft-azure/foundry/examples/deploy-vibevoice-asr}{\textbf{Microsoft Foundry}} \\
\end{tabular}
\end{table}

\begin{figure}[!h]
\centering
\includegraphics[width=\linewidth]{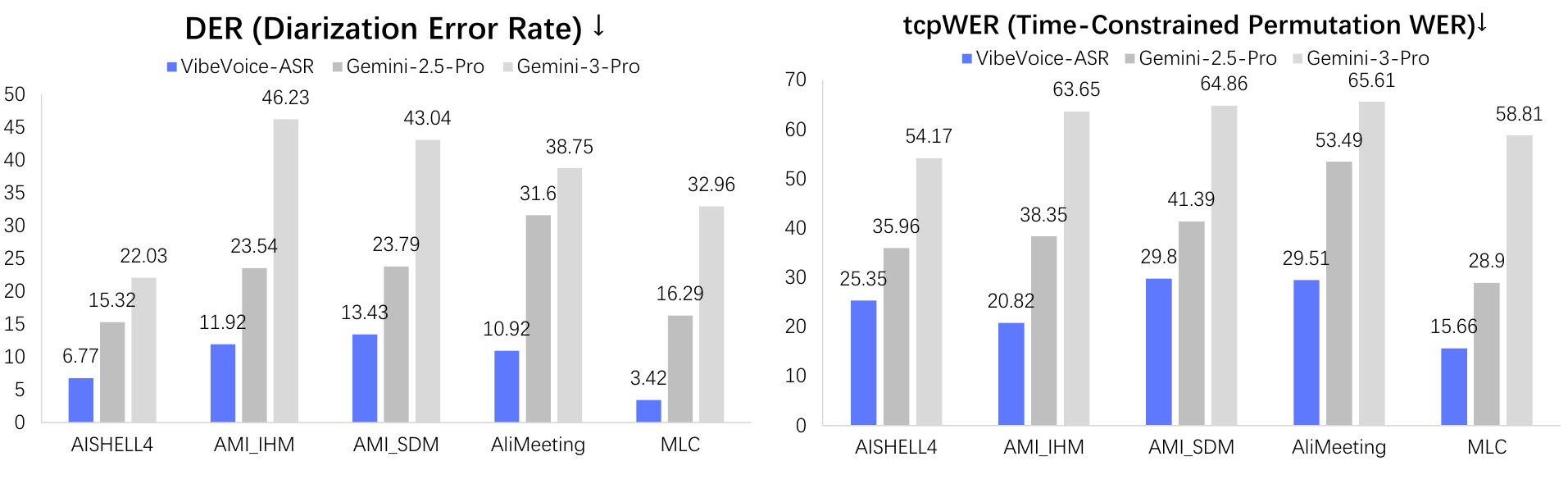} 
\caption{
% \our{} 
\our{} sets a new state-of-the-art for long-form speech understanding, consistently outperforming strong closed-source multimodal models (Gemini-2.5/3-Pro) across five public benchmarks. The results demonstrate superior accuracy in both speaker attribution (DER) and time-aligned transcription (tcpWER), particularly in complex multi-speaker environments.
}
\label{fig:result}
\end{figure}

% \newpage

\section{Introduction}

Recent years have witnessed a paradigm shift in speech processing, driven by the integration of Large Language Models (LLMs) with acoustic encoders~\cite{chu2023qwenaudio}. While these large audio models have achieved remarkable success in short-form speech recognition, transcribing and analyzing long-form audio—such as hour-long meetings, podcasts, and academic lectures—remains a formidable challenge.

The prevailing approach to long-form audio involves \textit{cascaded pipelines} that segment continuous speech into short clips (typically $<30$ seconds) for independent processing~\cite{he2024emilia, bain2023whisperx, bredin2020pyannote}. While practical, this "divide-and-conquer" strategy suffers from two fundamental limitations: Context Fragmentation and Pipeline Complexity. 
First, independently processing segments severs global semantic dependencies, causing the model to lose track of cross-sentence context, which is fatal for disambiguating homophones or resolving coreferences in extended dialogue. 
Second, traditional systems treat Automatic Speech Recognition (ASR), Speaker Diarization, and Timestamping as separate tasks managed by disjoint models. Reconciling their outputs often requires complex heuristics, leading to error propagation where a failure in segmentation or diarization corrupts the final transcript.

To bridge this gap, we introduce \our{}, a unified, general-purpose framework designed for high-fidelity long-form speech understanding. Built upon the VibeVoice architecture~\cite{vibevoice}, our system fundamentally abandons the sliding-window paradigm in favor of a single-pass approach. By leveraging an ultra-low frame rate tokenizer ($7.5$\,Hz), \our{} compresses an hour of audio into a sequence length that fits comfortably within the context window of modern LLMs. This allows the model to attend to the entire global context of a 60-minute session simultaneously, ensuring semantic coherence and consistent speaker tracking without the need for external clustering algorithms.
Concurrent with the development of \our{}, a number of related research efforts have emerged~\cite{huo2026tagspeech, yin2025speakerlm, shi2025train, yu2026moss}. Nevertheless, the majority of these works have not made their models publicly available.

\begin{figure}[t]
\centering
\includegraphics[width=1\linewidth]{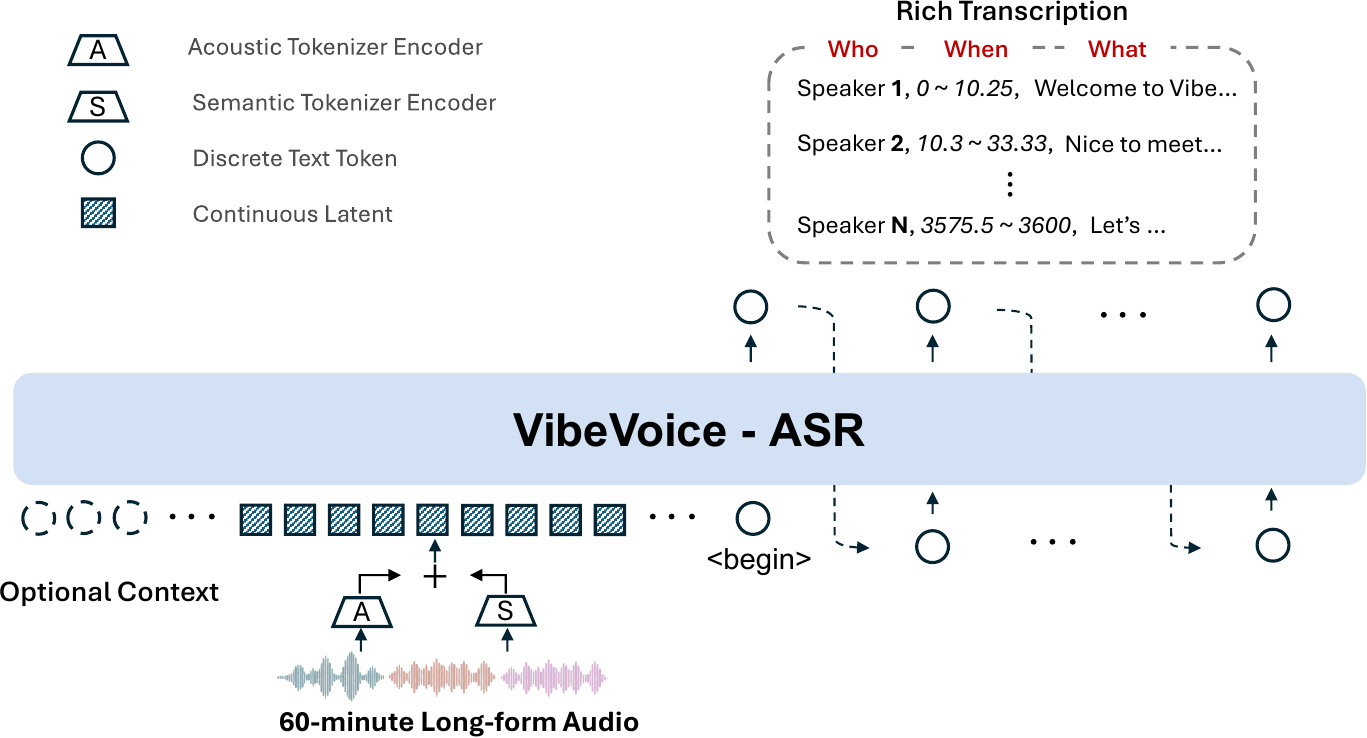}
\caption{
% \our{} 
The architectural overview of \our{}. \our{} processes 60-minute long-form audio in a single pass by ingesting continuous latents from dual-tokenizers alongside optional user-provided context. The output is a generated stream of Rich Transcription, explicitly interleaving Speaker ID (Who), Timestamps (When), and Content (What)
}
\label{fig:overall}
\end{figure}

\our{} reformulates long-form transcription as an end-to-end generation task, as shown in Figure~\ref{fig:overall}. Instead of outputting plain text, it generates a structured \textit{Rich Transcription} stream that explicitly interleaves speaker identities (``Who"), precise timestamps (``When"), and speech content (``What"). Furthermore, acknowledging the diverse needs of real-world applications, we introduce a prompt-based context injection mechanism. This allows users to supply customized context—ranging from hotword lists to background descriptions—significantly enhancing the model's ability to recognize domain-specific terminology and handle complex code-switching scenarios.

\section{Method}

\subsection{Overview}
Figure~\ref{fig:overall} presents the architectural overview of \our{}. We formulate long-form speech understanding as a language modeling task. The model takes a sequence of continuous audio embeddings, encoded from from the pre-trained Acoustic and Semantic encoders, as its primary input. To enable context-aware capabilities, optional text prompts (e.g., hotwords or background information) can be prepended to the audio sequence. 

These inputs are processed by a decoder-only Large Language Model backbone (e.g., Qwen 2.5~\cite{qwen2_5}) to autoregressively generate the target sequence. Distinct from conventional ASR models that output plain text, \our{} is designed to produce a \textit{Rich Transcription}. As illustrated in the output stream of Figure~\ref{fig:overall}, the model generates a structured sequence that explicitly interleaves speaker identity (``Who''), temporal boundaries (``When''), and speech content (``What''), enabling simultaneous recognition, diarization, and timestamping in a single pass.

\subsection{Speech Tokenizer}

In this work, we directly employ the pre-trained dual-tokenizers from \ourtts{}~\cite{vibevoice}, which integrates an Acoustic Tokenizer for spectral fidelity and a Semantic Tokenizer for linguistic alignment. The Acoustic tokenizer, inspired by \sigvae{}~\cite{latentlm}, applies a hierarchical design with a cumulative $3200\times$ downsampling rate to the $24$\,kHz input, yielding an extremely compact representation of approximately $7.5$ tokens per second. Meanwhile, the Semantic module extracts deterministic content features aligned with textual semantics. 
Note we only use tokenizer encoders here.
This ultra-low frame rate is pivotal, as a one-hour continuous audio session translates to:
\begin{equation}
    3600 \text{ seconds} \times 7.5 \text{ tokens/sec} = 27,000 \text{ tokens},
\end{equation}
which fits comfortably within the single-pass context window of modern LLMs.
 
\subsection{\our{}}

\subsubsection{Pre-training}

We use the data processing pipeline proposed in \ourtts{}~\cite{vibevoice, yu2024autoprep} to obtain the initial data corpus. 
The pre-training data distribution can be found in Figure~\ref{fig:data_distribution}.
The pipeline consists of three stages: segmentation and transcription, diarization, and quality filtering. 
Long recordings are first segmented using Silero voice activity detection (VAD) into clips of up to 30 seconds, followed by transcription with Whisper-large-v3-turbo~\citep{whisper} to obtain punctuated text and word-level timestamps; segment boundaries are further refined by splitting at punctuation end timestamps (e.g., \texttt{[.?!]}) to better align with speaker turns. 
Speech diarization is then performed using the vblinkp model from the WeSpeaker toolkit~\citep{wang2023wespeaker}, where speaker embeddings are extracted from overlapping frames (1.5 s window, 0.75 s hop), clustered with HDBSCAN~\citep{campello2013density}, and refined by merging clusters whose centroids have a cosine similarity greater than 0.67, yielding final speaker turn annotations. 
Finally, to ensure annotation reliability, segments are re-transcribed using a secondary ASR model~\citep{xu2023nemo}, and recordings are discarded if more than 30\% of segments have a WER exceeding 20\%, if speech accounts for less than 60\% of the total duration.

To ensure the effectiveness of the data processing pipeline, we conducted a comparative study between our pipeline and two widely adopted audio processing pipelines, WhisperX~\cite{bain2023whisperx} and Emilia~\cite{he2024emilia}. The evaluation is performed on three commonly used public multi-speaker meeting datasets—AMI~\cite{carletta2005ami}, AliMeeting~\cite{yu2022m2met}, and AISHELL-4~\cite{fu2021aishell}—and reports both diarization error rate (DER) and diarization invariant word error rate (WER). For a fair comparison, we disable the data-filtering module in Emilia, as its default configuration removes a substantial portion of the audio samples.

\begin{table}[h]
\centering
% \small
\caption{DER and WER comparison across different data pipelines.}
\begin{tabular}{l|cc|cc|cc|cc}
\toprule
Model & 
\multicolumn{2}{c|}{AISHELL4} &
\multicolumn{2}{c|}{AMI-IHM} &
\multicolumn{2}{c|}{AMI-SDM} &
\multicolumn{2}{c}{AliMeeting} \\
 & DER & WER & DER & WER & DER & WER & DER & WER \\
\midrule
WhisperX &
\textbf{14.55} & 29.69 &
18.27 & 24.12 &
23.05 & 39.65 &
35.53 & 36.62 \\
Emilia &
16.58 & 49.40 &
35.44 & 47.85 &
46.55 & 61.70 &
25.57 & 54.27  \\
Ours pipeline &
16.93 & \textbf{18.99} &
\textbf{15.46} & \textbf{23.22} &
\textbf{17.78} & \textbf{28.40} &
\textbf{25.34} & \textbf{30.82} \\
\bottomrule
\end{tabular}
\label{tab:der_wer}
\end{table}

As shown in Table~\ref{tab:der_wer}, the proposed data pipeline consistently achieves lower DER and WER than both baseline systems across the majority of evaluated datasets. These results indicate that our pipeline provides more robust segmentation, diarization, and transcription performance under diverse acoustic conditions.

We employed a curriculum learning strategy for the LLM input sequence
length, progressively increasing from 8,192 to 65,536 tokens. 
% Training \our{} took approximately 6 days on 128 AMD Instinct MI300X GPUs.

\subsubsection{Supervised Fine-Tuning (SFT)}

Since the pre-training stage predominantly relies on pseudo-labeled data, the SFT phase is critical for aligning the model with precise instruction-following behaviors. We carefully curate a high-quality dataset composition strategy, categorized into three distinct sources:

\paragraph{High-Quality Speech and Music Benchmarks.}
To establish a robust baseline for conversational speech recognition and speaker diarization, we utilize established datasets including the training splits of MLC-SLM~\cite{mlc} and  Fisher~\cite{cieri2004fisher}. These provide high quality labels for multi-speaker interactions. Additionally, we incorporate the open-source synthesized music dataset Muse~\cite{jiang2026muse} as an independent subset. The inclusion of this music data allows the model to learn music-specific acoustic features, explicitly optimizing its performance and robustness when handling musical segments.

\paragraph{Context-Aware Synthetic Data Pipeline.}
A key capability of \our{} is utilizing user-provided \textit{Contextual Information}—ranging from specific entities to complete sentences and background descriptions—to guide recognition. To bridge the lack of such paired data in real-world scenarios, we constructed a synthetic pipeline:
\begin{itemize}
    \item \textit{Context-Driven Script Generation:} We employ GPT-5~\cite{gpt5} to generate complex dialogue scripts containing specific entities, technical terms, and cross-lingual content (English, Chinese, and intra-sentential code-switching). Crucially, GPT-5 simultaneously generates the corresponding \textit{contextual reference text} (e.g., keyword lists, related sentences, or background paragraphs) used to prompt the ASR model.
    \item \textit{Audio Synthesis:} We leverage the \ourtts{} engine to synthesize high-fidelity multi-speaker audio. The synthesis predominantly targets Chinese, English, and complex English-Chinese code-switching scenarios, fully exploiting \ourtts{}'s superior capabilities in modeling these specific linguistic distributions and transitions.
    \item \textit{Quality Filtering:} We perform a closed-loop verification where the synthesized speech is transcribed back; samples exceeding a WER threshold are discarded to prevent noise injection. After that, we obtains about 6,000 hours synthesized audio.
\end{itemize}

\paragraph{Long-Form Transcription Restoration.}
Existing high-quality datasets are predominantly short ($<$30 minutes), creating a distribution shift for long-form applications. While we recall long-duration samples ($>$50 minutes) from our pre-training corpus, their original transcriptions—derived from our chunk-wise pipelines—also suffer from context fragmentation. To address this, we employ GPT-5 as a text refiner to rewrite and merge disjointed transcriptions into coherent, globally consistent long texts ("Global Semantic Rectification"). 
% To mitigate hallucinations, only rewritten transcripts that maintain high alignment with the original audio are retained.

Furthermore, to handle the non-speech intervals inherent in long-duration recordings, we utilize GPT-Audio\footnote{\url{https://platform.openai.com/docs/models/gpt-4o-audio-preview}} to automatically annotate these segments with general acoustic tags. Specifically, we label events such as \texttt{[Unintelligible Speech]}, \texttt{[Music]}, \texttt{[Human Sounds]}, \texttt{[Environmental Sounds]}, \texttt{[Noise]}, and \texttt{[Silence]}. This explicit tagging strategy provides direct supervision for non-speech intervals, designed to prevent the model from hallucinating text during silence or background noise.

To balance the \our{}'s capabilities across standard recognition, music robustness, context awareness, and long-form coherence, we apply a strategic data mixing ratio. Specifically, the sampling weights for Standard Benchmarks, Music Data, Synthetic Data, and Refined Long-Form Data are set to $0.5 : 0.1 : 0.1 : 0.3$, respectively. 

% The SFT phase is highly efficient: the entire process completes in approximately $3$ hours on $16$ AMD Instinct MI300X GPUs.

\section{Results}

We follow the MeetEval\footnote{\url{https://github.com/fgnt/meeteval}} evaluation protocol and report four complementary metrics that capture different aspects of multi-speaker transcription quality. 

\textbf{Diarization Error Rate (DER)} measures the accuracy of speaker attribution by accounting for speaker confusion, missed speech, and false alarm speech, and thus directly evaluates the model’s ability to answer who speaks when. 

\textbf{Word Error Rate (WER)} ignores speaker labels and timing information and computes the standard word-level error rate over the entire transcription, serving as a measure of pure speech recognition accuracy (what) independent of diarization performance.

\textbf{Concatenated minimum-Permutation WER (cpWER)} evaluates transcription accuracy under speaker permutation invariance by concatenating all utterances belonging to the same speaker and computing the minimum WER over all possible speaker permutations; this metric jointly reflects content recognition accuracy and speaker consistency, while being insensitive to local time alignment errors. 

\textbf{Time-Constrained minimum-Permutation WER (tcpWER)} further extends cpWER by enforcing temporal alignment constraints, such that words are only matched if they occur within a predefined temporal collar, making tcpWER sensitive to both speaker attribution and word-level timing accuracy and thus jointly evaluating who, what, and when. 

We select Gemini-2.5-Pro and Gemini-3-Pro as comparison baselines, as they represent state-of-the-art large-scale multimodal foundation models capable of jointly predicting timestamps, speaker labels, and transcription content. During our experiments, we observe that Gemini models exhibit substantial timestamp inaccuracies and occasional content hallucinations when processing long-form audio inputs. To ensure a fair and stable comparison, we therefore segment the test audio into 240-second chunks before feeding them to the Gemini models. In contrast, \our{} processes the entire audio recording in a single pass, without requiring chunk-wise inference.

\begin{table}[htb]
\centering
\caption{Overall diarization and ASR results across datasets and languages.}
\label{tab:all_datasets_languages}
\scriptsize
\setlength{\tabcolsep}{3pt}
\begin{tabular}{ll|cccc|cccc|cccc}
\toprule
 &  & \multicolumn{4}{c}{Gemini-2.5-Pro} 
 & \multicolumn{4}{c}{Gemini-3-Pro} 
 & \multicolumn{4}{c}{\our{}} \\
Dataset & Language 
& DER & cpWER & tcpWER & WER 
& DER & cpWER & tcpWER & WER 
& DER & cpWER & tcpWER & WER \\
\midrule
AISHELL-4  & Chinese
& 15.32 & 31.59 & 35.96 & 22.42
& 22.03 & 27.43 & 54.17 & 22.75
& \textbf{6.77} & \textbf{24.99} & \textbf{25.35} & \textbf{21.40} \\
\midrule
AMI-IHM    & English
& 23.54 & 29.57 & 38.35 & 18.48
& 46.23 & 22.34 & 63.65 & \textbf{17.61}
& \textbf{11.92} & \textbf{20.41} & \textbf{20.82} & 18.81 \\
AMI-SDM    & English
& 23.79 & 34.78 & 41.39 & 22.35
& 43.04 & \textbf{26.91} & 64.86 & \textbf{22.09}
& \textbf{13.43} & 28.82 & \textbf{29.80} & 24.65 \\
\midrule
AliMeeting & Chinese
& 31.60 & 41.64 & 53.49 & 27.43
& 38.75 & 32.84 & 65.61 & \textbf{26.75}
& \textbf{10.92} & \textbf{29.33} & \textbf{29.51} & 27.40 \\
\midrule
\multirow{12}{*}{MLC-Challenge}
 & English
 & 20.67 & 16.23 & 26.72 & 9.76
 & 30.88 & 12.85 & 57.64 & 10.19
 & \textbf{4.28} & \textbf{11.48} & \textbf{13.02} & \textbf{7.99} \\
 & French
 & 7.66 & 23.06 & 24.60 & 17.17
 & 40.82 & 22.02 & 71.11 & 18.71
 & \textbf{3.80} & \textbf{18.80} & \textbf{19.64} & \textbf{15.21} \\
 & German
 & 18.19 & 30.36 & 39.43 & 17.76
 & 42.14 & 23.56 & 73.86 & 19.39
 & \textbf{1.04} & \textbf{17.10} & \textbf{17.26} & \textbf{16.30} \\
 & Italian
 & 12.55 & 16.88 & 25.20 & \textbf{12.87}
 & 23.45 & \textbf{15.59} & 49.89 & 13.32
 & \textbf{2.08} & 15.76 & \textbf{15.91} & 13.91 \\
 & Japanese
 & 20.40 & 30.41 & 37.36 & 16.58
 & 59.68 & 21.96 & 81.41 & 18.47
 & \textbf{0.82} & \textbf{15.33} & \textbf{15.41} & \textbf{14.69} \\
 & Korean
 & 17.57 & 19.23 & 29.81 & 10.18
 & 39.28 & 19.39 & 57.33 & 11.21
 & \textbf{4.52} & \textbf{15.35} & \textbf{16.07} & \textbf{9.65} \\
 & Portuguese
 & 20.86 & 30.03 & 40.20 & 20.15
 & 39.17 & \textbf{23.29} & 85.44 & \textbf{20.10}
 & \textbf{7.98} & 29.91 & \textbf{31.65} & 21.54 \\
 & Russian
 & 5.35 & 14.26 & 16.59 & 10.74
 & 22.76 & 13.05 & 51.89 & \textbf{10.31}
 & \textbf{0.90} & \textbf{12.94} & \textbf{12.98} & 12.40 \\
 & Spanish
 & 9.10 & 13.82 & 17.49 & 9.09
 & 25.54 & 12.11 & 43.72 & 9.36
 & \textbf{2.67} & \textbf{10.51} & \textbf{11.71} & \textbf{8.04} \\
 & Thai
 & 15.54 & 20.84 & 30.28 & 14.84
 & 22.09 & \textbf{14.59} & 39.54 & \textbf{12.03}
 & \textbf{4.09} & 14.91 & \textbf{15.57} & 13.61 \\
 & Vietnamese
 & 14.65 & 16.71 & 27.28 & 12.33
 & 32.24 & \textbf{13.15} & 60.43 & \textbf{11.53}
 & \textbf{0.16} & 14.57 & \textbf{14.57} & 14.43 \\
 \cline{2-14}
 & AVERAGE
 & 16.29 & 20.37 & 28.90 & 13.05
 & 32.96 & 16.38 & 58.81 & 13.11
 & \textbf{3.42} & \textbf{14.81} & \textbf{15.66} & \textbf{12.07} \\
\bottomrule
\end{tabular}
\end{table}

As shown in Table~\ref{tab:all_datasets_languages},
\our{} consistently outperforms Gemini-2.5-Pro and Gemini-3-Pro in terms of DER and tcpWER across all evaluated datasets, demonstrating substantially stronger speaker modeling and more accurate alignment of speaker turns over time. 
On the cpWER metric, which more directly reflects the model’s ability to maintain speaker consistency, our model achieves the best performance on 11 out of 16 evaluation settings, significantly outperforming both Gemini variants and indicating more reliable speaker differentiation in multi-speaker conditions. 
Regarding WER, our model attains the lowest error rate on 8 out of 16 settings, while exhibiting only marginal degradation on the remaining datasets. 
Overall, these results indicate that \our{} achieves a better balance between content recognition accuracy and robust speaker-aware transcription, with particularly strong advantages in speaker attribution, temporal consistency, and multilingual generalization.

\section{Conclusion and Limitations}

In this report, we presented \our{}, a unified single-pass framework that effectively solves context fragmentation in long-form speech understanding. Beyond technical contributions, we commit to \textbf{comprehensive open-sourcing}, releasing the model weights, fine-tuning pipelines, and high-performance inference code (e.g., vLLM~\cite{vllm} support). By democratizing access to these tools, we aim to empower the research community to address the SFT gaps in low-resource languages and adapt the framework to diverse downstream applications, ultimately fostering a more inclusive and advanced speech ecosystem.

Despite these advancements, \our{} has several limitations that guide future research:
\begin{itemize}
    \item \textit{Multilingual Forgetting in SFT:} While our pre-training covered over 50 languages, the SFT phase predominantly focused on English, Chinese, and code-switching data. Consequently, the model may experience performance degradation on low-resource languages absent from the instruction tuning stage. We hope our open-source fine-tuning code will encourage the community to bridge this gap.
    \item \textit{Overlapping Speech:} The current architecture generates a serialized output stream and does not explicitly handle overlapping speech (the "cocktail party problem"). In scenarios where multiple speakers talk simultaneously, the model tends to transcribe the dominant speaker, potentially missing secondary information. Future iterations will explore separation-aware modeling to address this challenge.
\end{itemize}

\section*{Acknowledge}
We thank Ruibin Yuan, Tao Zhang and Zhengwei Huang for their in-depth discussions during the research and development of \our{}.

% \newpage
\bibliography{vibepod}

@article{latentlm,
  title={Multimodal Latent Language Modeling with Next-Token Diffusion},
  author={Sun, Yutao and Bao, Hangbo and Wang, Wenhui and Peng, Zhiliang and Dong, Li and Huang, Shaohan and Wang, Jianyong and Wei, Furu},
  journal={arXiv preprint arXiv:2412.08635},
  year={2024}
}

@article{qwen2_5,
  title={Qwen2. 5 technical report},
  author={Yang, An and Yang, Baosong and Zhang, Beichen and Hui, Binyuan and Zheng, Bo and Yu, Bowen and Li, Chengyuan and Liu, Dayiheng and Huang, Fei and Wei, Haoran and others},
  journal={arXiv preprint arXiv:2412.15115},
  year={2024}
}

@inproceedings{yu2024autoprep,
  title={Autoprep: An automatic preprocessing framework for in-the-wild speech data},
  author={Yu, Jianwei and Chen, Hangting and Bian, Yanyao and Li, Xiang and Luo, Yi and Tian, Jinchuan and Liu, Mengyang and Jiang, Jiayi and Wang, Shuai},
  booktitle={ICASSP 2024-2024 IEEE International Conference on Acoustics, Speech and Signal Processing (ICASSP)},
  pages={1136--1140},
  year={2024},
  organization={IEEE}
}

@inproceedings{he2024emilia,
  title={Emilia: An extensive, multilingual, and diverse speech dataset for large-scale speech generation},
  author={He, Haorui and Shang, Zengqiang and Wang, Chaoren and Li, Xuyuan and Gu, Yicheng and Hua, Hua and Liu, Liwei and Yang, Chen and Li, Jiaqi and Shi, Peiyang and others},
  booktitle={2024 IEEE Spoken Language Technology Workshop (SLT)},
  pages={885--890},
  year={2024},
  organization={IEEE}
}

@inproceedings{campello2013density,
  title={Density-based clustering based on hierarchical density estimates},
  author={Campello, Ricardo JGB and Moulavi, Davoud and Sander, J{\"o}rg},
  booktitle={Pacific-Asia conference on knowledge discovery and data mining},
  pages={160--172},
  year={2013},
  organization={Springer}
}

@inproceedings{whisper,
  title={Robust speech recognition via large-scale weak supervision},
  author={Radford, Alec and Kim, Jong Wook and Xu, Tao and Brockman, Greg and McLeavey, Christine and Sutskever, Ilya},
  booktitle={International conference on machine learning},
  pages={28492--28518},
  year={2023},
  organization={PMLR}
}

@inproceedings{xu2023nemo,
  title={Efficient sequence transduction by jointly predicting tokens and durations},
  author={Xu, Hainan and Jia, Fei and Majumdar, Somshubra and Huang, He and Watanabe, Shinji and Ginsburg, Boris},
  booktitle={International Conference on Machine Learning},
  pages={38462--38484},
  year={2023},
  organization={PMLR}
}

@inproceedings{wang2023wespeaker,
  title={Wespeaker: A research and production oriented speaker embedding learning toolkit},
  author={Wang, Hongji and Liang, Chengdong and Wang, Shuai and Chen, Zhengyang and Zhang, Binbin and Xiang, Xu and Deng, Yanlei and Qian, Yanmin},
  booktitle={ICASSP 2023-2023 IEEE International Conference on Acoustics, Speech and Signal Processing (ICASSP)},
  pages={1--5},
  year={2023},
  organization={IEEE}
}

@article{vibevoice,
  title={Vibevoice technical report},
  author={Peng, Zhiliang and Yu, Jianwei and Wang, Wenhui and Chang, Yaoyao and Sun, Yutao and Dong, Li and Zhu, Yi and Xu, Weijiang and Bao, Hangbo and Wang, Zehua and others},
  journal={arXiv preprint arXiv:2508.19205},
  year={2025}
}

@article{mlc,
  title={Summary on The Multilingual Conversational Speech Language Model Challenge: Datasets, Tasks, Baselines, and Methods},
  author={Mu, Bingshen and Guo, Pengcheng and Sun, Zhaokai and Wang, Shuai and Liu, Hexin and Shao, Mingchen and Xie, Lei and Chng, Eng Siong and Xiao, Longshuai and Feng, Qiangze and others},
  journal={arXiv preprint arXiv:2509.13785},
  year={2025}
}

@article{jiang2026muse,
  title={Muse: Towards Reproducible Long-Form Song Generation with Fine-Grained Style Control},
  author={Jiang, Changhao and Chen, Jiahao and Xiang, Zhenghao and Yang, Zhixiong and Wang, Hanchen and Zhuang, Jiabao and Che, Xinmeng and Sun, Jiajun and Li, Hui and Cao, Yifei and others},
  journal={arXiv preprint arXiv:2601.03973},
  year={2026}
}

@inproceedings{carletta2005ami,
  title={The AMI meeting corpus: A pre-announcement},
  author={Carletta, Jean and Ashby, Simone and Bourban, Sebastien and Flynn, Mike and Guillemot, Mael and Hain, Thomas and Kadlec, Jaroslav and Karaiskos, Vasilis and Kraaij, Wessel and Kronenthal, Melissa and others},
  booktitle={International workshop on machine learning for multimodal interaction},
  pages={28--39},
  year={2005},
  organization={Springer}
}

@article{fu2021aishell,
  title={Aishell-4: An open source dataset for speech enhancement, separation, recognition and speaker diarization in conference scenario},
  author={Fu, Yihui and Cheng, Luyao and Lv, Shubo and Jv, Yukai and Kong, Yuxiang and Chen, Zhuo and Hu, Yanxin and Xie, Lei and Wu, Jian and Bu, Hui and others},
  journal={arXiv preprint arXiv:2104.03603},
  year={2021}
}

@inproceedings{yu2022m2met,
  title={M2MeT: The ICASSP 2022 multi-channel multi-party meeting transcription challenge},
  author={Yu, Fan and Zhang, Shiliang and Fu, Yihui and Xie, Lei and Zheng, Siqi and Du, Zhihao and Huang, Weilong and Guo, Pengcheng and Yan, Zhijie and Ma, Bin and others},
  booktitle={ICASSP 2022-2022 IEEE International Conference on Acoustics, Speech and Signal Processing (ICASSP)},
  pages={6167--6171},
  year={2022},
  organization={IEEE}
}

@article{bain2023whisperx,
  title={Whisperx: Time-accurate speech transcription of long-form audio},
  author={Bain, Max and Huh, Jaesung and Han, Tengda and Zisserman, Andrew},
  journal={arXiv preprint arXiv:2303.00747},
  year={2023}
}

@article{chu2023qwenaudio,
  title={Qwen-audio: Advancing universal audio understanding via unified large-scale audio-language models},
  author={Chu, Yunfei and Xu, Jin and Zhou, Xiaohuan and Yang, Qian and Zhang, Shiliang and Yan, Zhijie and Zhou, Chang and Zhou, Jingren},
  journal={arXiv preprint arXiv:2311.07919},
  year={2023}
}

@inproceedings{vllm,
  title={Efficient Memory Management for Large Language Model Serving with PagedAttention},
  author={Woosuk Kwon and Zhuohan Li and Siyuan Zhuang and Ying Sheng and Lianmin Zheng and Cody Hao Yu and Joseph E. Gonzalez and Hao Zhang and Ion Stoica},
  booktitle={Proceedings of the ACM SIGOPS 29th Symposium on Operating Systems Principles},
  year={2023}
}

@inproceedings{cieri2004fisher,
  title={The Fisher corpus: A resource for the next generations of speech-to-text.},
  author={Cieri, Christopher and Miller, David and Walker, Kevin},
  booktitle={LREC},
  volume={4},
  pages={69--71},
  year={2004}
}

@article{gpt5,
  title={Openai gpt-5 system card},
  author={Singh, Aaditya and Fry, Adam and Perelman, Adam and Tart, Adam and Ganesh, Adi and El-Kishky, Ahmed and McLaughlin, Aidan and Low, Aiden and Ostrow, AJ and Ananthram, Akhila and others},
  journal={arXiv preprint arXiv:2601.03267},
  year={2025}
}

@inproceedings{bredin2020pyannote,
  title={Pyannote. audio: neural building blocks for speaker diarization},
  author={Bredin, Herv{\'e} and Yin, Ruiqing and Coria, Juan Manuel and Gelly, Gregory and Korshunov, Pavel and Lavechin, Marvin and Fustes, Diego and Titeux, Hadrien and Bouaziz, Wassim and Gill, Marie-Philippe},
  booktitle={ICASSP 2020-2020 IEEE International Conference on Acoustics, Speech and Signal Processing (ICASSP)},
  pages={7124--7128},
  year={2020},
  organization={IEEE}
}

@article{huo2026tagspeech,
  title={TagSpeech: End-to-End Multi-Speaker ASR and Diarization with Fine-Grained Temporal Grounding},
  author={Huo, Mingyue and Shao, Yiwen and Zhang, Yuheng},
  journal={arXiv preprint arXiv:2601.06896},
  year={2026}
}

@article{yin2025speakerlm,
  title={SpeakerLM: End-to-end versatile speaker diarization and recognition with multimodal large language models},
  author={Yin, Han and Chen, Yafeng and Deng, Chong and Cheng, Luyao and Wang, Hui and Tan, Chao-Hong and Chen, Qian and Wang, Wen and Li, Xiangang},
  journal={arXiv preprint arXiv:2508.06372},
  year={2025}
}

@article{shi2025train,
  title={Train Short, Infer Long: Speech-LLM Enables Zero-Shot Streamable Joint ASR and Diarization on Long Audio},
  author={Shi, Mohan and Xiao, Xiong and Fan, Ruchao and Ling, Shaoshi and Li, Jinyu},
  journal={arXiv preprint arXiv:2511.16046},
  year={2025}
}

@article{yu2026moss,
  title={MOSS Transcribe Diarize: Accurate Transcription with Speaker Diarization},
  author={Yu, Donghua and Lin, Zhengyuan and Yang, Chen and Zhang, Yiyang and Fei, Zhaoye and Chen, Hanfu and Chen, Jingqi and Chen, Ke and Cheng, Qinyuan and Fan, Liwei and others},
  journal={arXiv preprint arXiv:2601.01554},
  year={2026}
}
\bibliographystyle{alpha}

\clearpage
\appendix
\section{Language Distribution of Training Data }

\begin{figure}[!bh]
\centering
\includegraphics[
  width=1\linewidth,
  % height=0.853\textheight,
  % height=1.0\textheight,
  keepaspectratio
]{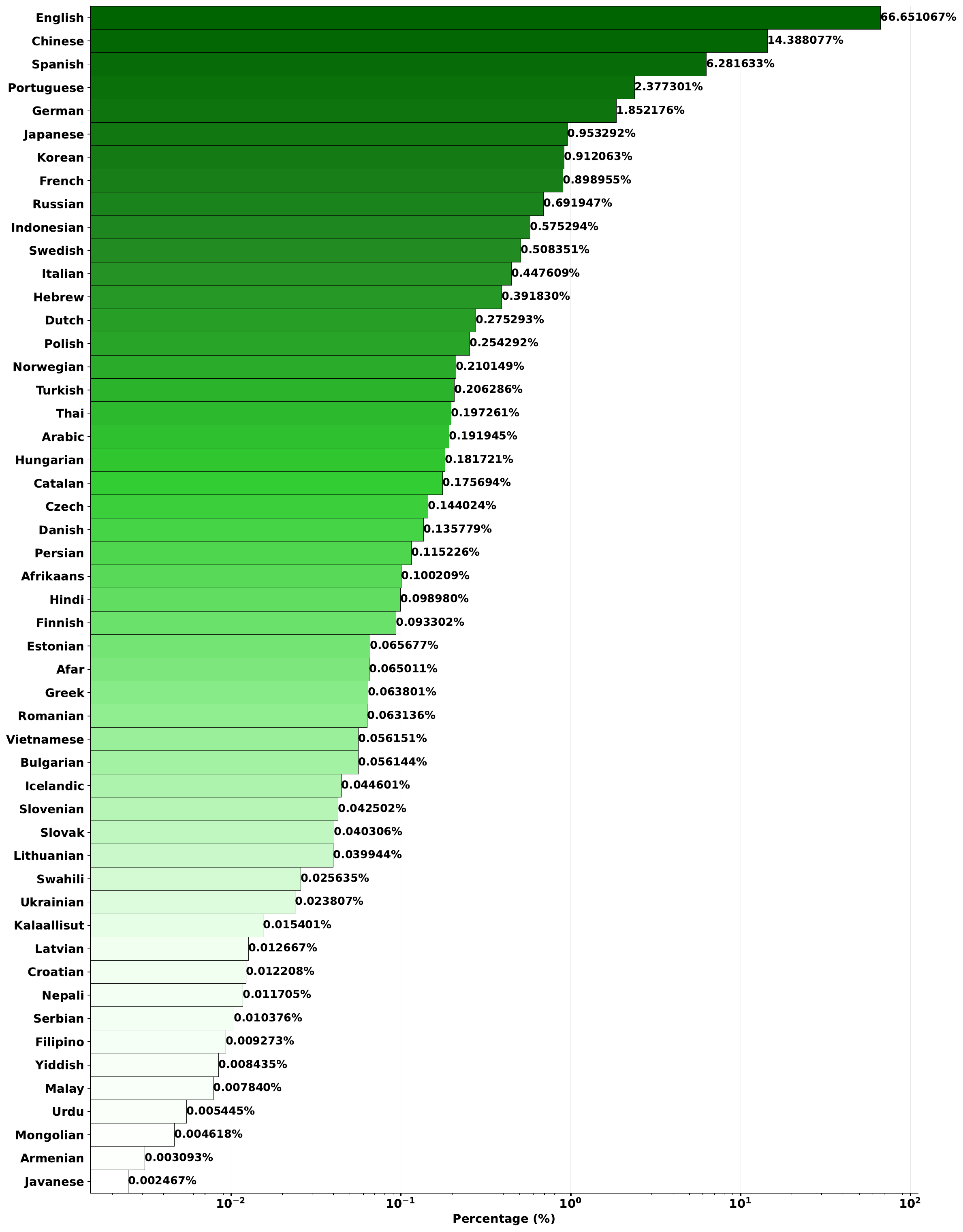}
\caption{
Language distribution in the training data.
}
\label{fig:data_distribution}
\end{figure}

\end{document}